# The influence of the network topology on the agility of a supply chain

**Hernández, Juan M.**[*]
Department of Quantitative Methods, Institute of Tourism and Sustainable Economic Development (TIDES), University of Las Palmas de Gran Canaria, c/ Saulo Torón, nº17, 35017 Spain. E-mail: juan.hernandez@ulpgc.es
*Corresponding author

**Pedroza, Carmen**
Unidad Académica de Estudios Regionales de la Coordinación de Humanidades de la Universidad Nacional Autónoma de México (UNAM), Sede La Ciénega, Av. Lázaro Cárdenas s/n, esq. Felícitas del Río, Colonia, Centro, C.P. 59510, Jiquilpan de Juárez, Michoacán, Mexico. E-mail: pedrozacarmen@yahoo.es

## Abstract.

The right performance of a supply chain depends on the pattern of relationships among firms. Although there is not a general consensus among researchers yet, many studies point that scale-free topologies, where few highly related firms are combined with many low-related firms, assure the highest efficiency of a supply chain. This paper studies the network topology that leads to the highest agility of the supply chain when sudden demand changes occur. To do this, an agent-based model of a supply chain with restricted relationship between agents is built. The model includes three tiers, where the flow of material is distributed from the bottom supplier to the final customer passing necessarily through firms in every tier. Agility is measured in the model simulations through the order fulfillment rate. Unlike to previous theoretical and lab results, the simulation of the model shows that the highest levels of agility are not obtained with a scale-free topology. Instead, homogeneous distribution of links, such as those induced by regular or Poisson probability laws, shows higher agility values than heterogeneous distributions. Other previous recommendations, such as redundancy or having multiple suppliers, are confirmed by the simulations. The general conclusion is that the most suitable network topology in terms of agility depends on the specific conditions of the supply chain and the aspects of the performance to be analyzed.

**Keywords:** topology, agility, supply network, scale-free, homogeneous distribution

## 1. Introduction

Supply chain (SC) has been conceptualized as a network of agents (e.g., suppliers, manufacturers, distributors, retailers and consumers) who are interconnected throughout the transference of material or information (Carter et al., 2015). This perspective departs from the dyadic (buyer-supplier) and triadic view which center on the particular relationships between categories (Choi and Wu, 2009; Cox et al., 2001) to a more extended focus where the object of analysis is the supply chain as a whole, represented by a collection of nodes (agents) and links (transactions among them). Assuming this view, the social network analysis is an adequate tool to analyze this kind of systems, making use of quantitative metrics to identify some key nodes or the structure of the network as a whole (Borgatti and Li, 2009).

Following the network perspective, a supply chain is also termed supply network (Choi and Hong, 2002; Kim et al., 2011). A supply network is represented by a complex system of interdependence among multiple firms, each one with a number of links to specific partners. In this context, the supply network performing is not just the result of an additive aggregation of the individual firms' decisions, such as it would be the case in a simple and linear SC. Instead of this, a coherent and autonomous global behavior emerges from the non-linear interaction among agents (Surana et al., 2005; Choi et al., 2001). This behavior is dependent on the particular structure of the total relationships among agents in the supply network, what is called the network topology. Although some scholars have made efforts to distinguish supply chains and supply networks recently (Braziotis et al., 2013), in this paper we use both terms interchangeably.

Some SC archetypes or topologies have been identified in the scientific literature (Capaldo and Giannoccaro, 2015a; Hearnshaw and Wilson, 2013; Kim et al., 2015). Among them, structures characterized by a heterogeneous distribution of links, such as the scale-free topology, are among the most frequently analyzed. Induced by power-law degree distribution, scale-free supply networks are characterized by the existence of few large hub firms with many connections combined with many small peripheral firms with few connections. This topology is presented in some real SCs and is seen by some authors as the most efficient topology in SC (Hearnshaw and Wilson, 2013). In particular, it has been theoretically shown that scale-free is the most suitable structure to respond against disturbances and disruptions, what determines a resilient SC (Kim et al., 2015; Nair and Vidal, 2011; Zhao et al., 2011).

In this paper, we analyze the influence of the network topology on the agility of the supply chain. SC agility can be defined as the ability to respond quickly to sudden changes in supply and demand and is considered as one of the dimensions of a resilient SC (Christopher and Peck, 2004). Agility is usually analyzed from a focal agent's view, assuming the position of a single company in the network and looking the surrounding supply chain in relation to this company (Lee, 2004). Unlikely, our focus adopts a more extended view and analyses agility of the network as a whole, not from the perspective of a single agent (Choi et al., 2001).

In supply networks, agility is influenced by the number of intermediate agents through which the flow of material or information needs to pass between the initial supplier and the final customer. In general, the theoretical analyses of resilience in SC so far assume that a link is possible between any two firms in the supply network, independently on the specific tier where the firms belong to (Nair and Vidal, 2011; Zhao et al., 2011). However, in some specific real supply networks the flow needs to pass through firms which belong to stablished tiers and bypass seldom occurs. This is the case for military logistic networks and many food supply chains. The main hypothesis of this paper is that, in these kinds of supply networks, topologies characterized by heterogeneous distributions of links, such as those induced by the power-law degree distribution, present lower values of agility than others induced by homogeneous distributions of links. Therefore, the paper limits to specific classes of SCs the extended belief that scale-free and derivatives are the most efficient SC topologies.

In order to check the hypotheses, we build an agent-based model that represents a supply chain with multiple agents and their rules to allocate orders and supplies. The SC in the model includes three tiers (suppliers, wholesalers and distributors) and the pattern of relationships among agents in different tiers follows some probabilistic distributions. Thus, the theoretical SC is built through the realization of random network models in bipartite graphs (Guillaume and Latapy, 2006). Several patterns of relationships between tiers are considered. Following the common recommendation of using agent-based simulation to analyze large, stochastic and non-centralized supply networks (Datta et al., 2007; Kim, 2009), we simulate a sudden demand change and measure the degree of success in adapting to these changes. The results show that, under certain conditions, the percentage of unfilled orders in the SC when assuming power-law degree distributions is substantially higher in value than when assuming other homogeneous distributions.

The paper is organized as follows. Next section presents a literature review, starting from which the hypotheses are stated. Section 3 shows the supply chain random model, order and supplies allocation rules, and the algorithm to measure SC agility. Section 4 presents the results of the simulations. Last two sections are devoted for discussion and conclusions.

## 2. Literature review

*2.1. Characteristics of the food supply chain*

In this subsection, we review some features of the food supply chain (FSC). One of them is the role of intermediaries, what makes that relationships among agents in a FSC are usually more restrictive than in industrial SCs. This paper analyzes agility in SC with restrictive relationships and FSCs are real examples proximate to these types of SCs.

FSCs are characterized by the perishability of their products. Additionally, many of their traded goods come from small-scale producers from rural economies in developing countries. Small-scale producers in the agricultural or fisheries sectors have also in common close relations with middlemen or intermediaries. To be able to sell their products, small-scale rural producers participate in a multi-tiered SC, where middlemen are key actors that connect producers with external markets. More specifically, middlemen act as facilitators of trade between producers and wholesalers. To give some examples, this market structure has been identified in many rural economies in agriculture and fishing, such as the potato market in Ethiopia (Abebe et al., 2016), the rice in Philipines (Hayami et al., 1999), or the fisheries Malaysia (Merlijn, 1989), Kenia (Crona et al., 2010) and Mexico (Pedroza, 2013).

Small-scale producers in these economies have low production volumes, limited trading skills and lack of conservation or transportation facilities. Furthermore, the lack of trading skills is usually related to low educational levels and isolated geographic conditions. Under these circumstances, the bulking and perishability of products become the main problems (Abebe et al., 2016; Keys, 2005). These factors do not help the small-scale producers to organize and have fluid horizontal relationships among them, which might allow them achieving local, national or international markets. An empirical evidence of the lack of strong horizontal relationships at producer levels in this type of economies is the malfunctioning of cooperatives. These producer organizations have served to organize production and obtain subsidies from government and foreign aid agencies, but are ineffective to improve commercialization and competitiveness (Abebe et al., 2016; Bernard and Taffesse, 2008; Hayami et al., 1999; Pedroza, 2013; Ruben and Heras, 2012).

The reasons why small-scale producers trade their products through middlemen and the influence that these intermediaries have in the FSC are from social and economic nature. Middlemen are a social optimal choice to decrease transactions costs, specifically those related to search and matching costs (Abebe et al., 2016; Arya et al., 2015; Gabre-Madhin, 2001). Middlemen act as a time saving institution by shortening the negotiation time of sellers and buyers in a transaction (Rubinstein and Wolinsky, 1987). Moreover, they are considered important providers of market information (Abebe et al., 2016), match geographically dispersed buyers and sellers (Gabre-Madhin, 2001), have better information and in general have the ability to contract separately with upstream and downstream parties and achieve supply and delivery coordination, in sum higher expertise (Arya et al., 2015). Middlemen link actors across different social domains and hierarchical levels (Crona et al., 2010), reducing risk and assuring fast distribution of food. Middlemen operate most of the time informally and through cash transactions. All these abilities are a source of flexibility, which enhances SC agility.

Middlemen go to rural areas to look for the product by means of a face-to-face relationship with each producer. Each small producer sells only a small amount, which tends to increase the transaction cost per unit of product (Hayami et al., 1999). Thus, middlemen can assure the demanded supply through its own network of producers. This represents a competitive advantage with respect to wholesalers who buy the product directly to few producers, which normally do not guarantee them sufficient supply (Abebe et al., 2016; Keys, 2005). Middlemen income level and relevance in the FSC depends on its capacity to increase, maintain and manage their collection volume, and this is normally done through their linking abilities. Commonly, middlemen in these markets have links with a large network of producers and a small network of wholesalers.

Therefore, middlemen play a key role in FSCs, since they can channel market demands and directly influence on the organization of production (Wilson, 1980). In a supply network where the product passes throughout the middlemen' hands, food will arrive on time and with a good quality to the table of consumers. This is the reason why middlemen are a necessary agent to transport the flow of product in a FSC.

*2.2. Agility in a resilient SC*

Agility has been recognized as one of the key qualities for top-performing supply chains (Lee, 2004). Additionally, it has been also identified as a constituent element of a resilient SC (Christopher and Peck, 2004; Jüttner and Maklan, 2011; Ponomarov and Holcomb, 2009; Wieland and Wallenburg, 2013). In this context, the concept of SC resilience embraces the one of agility, which is considered a dimension of it.

The definition and practice of supply chain resilience has been largely debated in the scientific literature since the beginning of the century. In general, resilience alludes to the SC capability to respond to unexpected events. Kamalahmadi and Parast (2016), based on a review of the different definitions given in the literature, propose the following extended definition: "The adaptive capability of a supply chain to reduce the probability of facing sudden disturbances, resist the spread of disturbances by maintaining control over structures and functions, and recover and respond by immediate and effective reactive plans to transcend the disturbance and restore the supply chain to a robust state of operations". This definition integrates agility in a resilient SC by mentioning explicitly the need of fast response to disturbances.

The different conceptualizations given since the beginning of the century have generally considered agility as a constituent (or formative) element of the SC resilience. Christopher and Peck (2004) considers it as a principle of resilience in combination with other three: supply chain re-engineering or the SC design, collaboration among entities in the SC and supply chain risk management culture, the last one favored by leadership and innovation. Other authors propose new conceptual frameworks to the question of resilience, but maintain a principal role for agility. For example, Pettit et al. (2010) enumerate 14 capabilities, divided in 111 subfactors, to counteract SC vulnerabilities, where agility is not explicitly included but split in several capability factors. Wieland and Wallenburg (2013) assume resilience divided between proactive (agility) and reactive (robustness) responses to disturbances. In empirical studies, agility was found as the major enabler of resilience by a sample of firm managers in Indian firms, followed by collaboration and visibility (Soni et al., 2014).

Thus, the definition and role of agility is repeated with few variations in most of the conceptualizations of resilience in the literature. Here we follow Christopher and Peck (2004), which defines agility as the "ability to respond rapidly to unpredictable changes in demand or supply". It is generally accepted that agility includes two dimensions: (a) visibility, or the "knowledge of the status of operating assets and the environment" (Pettit et al., 2010), which refers to the ability to see all entities, production and distribution capacities, together to other restrictions down and upstream in the SC, in order to avoid ineffective actions when a disturbance occurs, and (b) velocity, defined as the "the total time it takes to move product and materials from one end of the supply to the other" (Christopher and Peck, 2004), refers to the speed of the response when the change in demand and supply is produced.

Some authors also include flexibility as other ingredient of agility (Scholten et al., 2014; Jüttner and Maklan, 2011). We adopt the definitions of flexibility given by Pettit et al. (2010) as "ability to quickly change inputs or the mode of receiving inputs" (flexibility in sourcing) and the "ability to quickly change outputs or the mode of delivering outputs" (flexibility in order fulfillment). According to Sheffi and Rice (2005), flexibility can be achieved by working with multiple suppliers, having multiple capabilities at each plant location and a "coherent process for setting priorities during the time-sensitive postdisruption period". These strategies are related to the so-called redundancy, which refers to the existence of extra stock in case of disturbances (high inventory levels) or having multiple suppliers as well. Redundancy can be effective to get an agile response to disturbances although it uses to be costly (Sheffi and Rice, 2005).

So, agility is a multidimensional concept and influenced by multiple factors. Specifically, collaboration among the different entities in the SC exerts a positive influence on agility. The collaborative partners share information or material and "demonstrates a sense of responsibility towards its supply chain" (Wieland and Wallenburg, 2013). Thus, they can give a faster and more efficient response to disturbances by sharing resources. Collaboration can be vertical, if produced among agents in different tiers of the supply chain, or horizontal, if produced among entities in the same tier. Several empirical tests have proven the essential role of horizontal and vertical collaboration in SCs for an efficient disaster management process (Scholten et al., 2014). Additionally, the positive influence of collaboration on agility has been empirically tested by some recent studies (Scholten and Schilder, 2015; Wieland and Wallenburg, 2013).

In this paper we restrict the analysis to the relationship between the supply network topology and agility, leaving aside other factors that influence on it. We also depart from the focal agent's view and analyze agility of the supply network as a whole, not from the perspective of a single agent. In this context, we question the conditions for the supply network to be agile.

*2.3. The influence of topology on SC performance*

As commented in the introduction, we follow the network perspective to analyze SC agility (Borgatti and Li, 2009). Generally speaking, a network is a set of vertices (nodes), edges (links) and the way both elements are connected. In the context of SC, nodes are represented by firms, while edges represent relationships among those agents. This relationship can be formalized through the existence of some flow of material, information or contract among the parts (Beamon, 1998; Kim et al., 2011). For simplicity, in this paper we assume the same network structure for the three types of relationship. So, a link between firms indicates flow of material, information or the presence of a formal (or informal) agreement among them.

Supply networks have been also identified as complex adaptive systems (CAS) (Choi et al., 2001; Pathak et al., 2007; Surana et al., 2005). One of the most prominent characteristics of CAS is the presence of self-organization. This phenomenon occurs when a global functioning of the network emerges from the individual decisions of the partners which are combined in a non-linear and decentralized way. In a supply chain like this, the network behaves coherently from the aggregation of every agent's decisions, where everyone influences on the functioning of the network but not anyone controls the global behavior. In brief words, this can be summarized by the sentence "the whole is more than the sum of the parts".

The self-organization phenomenon in real networks has been also observed in other fields of science, such as Physics or Biology, and has given origin to the complex network theory (Albert and Barabasi, 2002; M. E. J. Newman, 2003). This theory analyses the global behavior of high-scale networks, such as is the case of Internet, author's citation and on-line recommendations, starting from simple rules of relationships among nodes. Some interdependence patterns or topologies of the network have been identified and their influence on action performance or dynamical models in the network, such as removing nodes or contagion diffusion, has been analyzed (Newman, 2010).

In real social networks, three common topological features have been regularly observed (Newman and Watts, 2002): (a) Evidences of "Small World"; (b) High clustering; and (c) Power-law degree distribution. Below we detail an explanation of these evidences and the corresponding facts in SCs. Larger formalization can be found in the review by Newman (2003).

The concept of "Small World" property has become popular from the experiment by Travers and Milgram (1969), who found that two strangers can be connected throughout few paths of acquaintances. In social network words, "Small World" indicates that the network diameter, defined as the maximum of the shortest path length among any two nodes in the network, is low as compared with the network size. In a SC context, this means that, although the supply chain includes a lot of firms and relationships, the flow of material or information passes through few firms from the initial supplier to the final customer.

Clustering in a network refers to the probability that a triadic relationship forms a closed cycle. In terms of a friendship relationship, this means how often two persons are friend if both are friends of a third one. Hence, the clustering coefficient of the network is the percentage of cycles in all the possible triadic relationships in a network. As expected, high clustering coefficient has been observed in many real social networks (Newman, 2003). In the supply chain context, triads can be formed by buyer-supplier-supplier relationships (Choi and Wu, 2009), but also by buyer-buyer-supplier and three firms in the same tier. High clustering has been observed in some real SCs (Hearnshaw and Wilson, 2013).

In network theory, the number of links from a node is called node degree. The network degree distribution ($p(k)$) indicates the proportion of nodes that have a *k* degree. In other words, the degree distribution indicates the probability that an arbitrary node includes *k* edges. Many real networks follow a power-law distribution, $p(k) \sim k^{-\gamma}$, with $\gamma$ a parameter usually between 1 and 3. One of the most outstanding characteristic of the power-law is the existence of fat tails in the probability distribution graph. In other words, in these networks there is few but significant amount of highly connected nodes together with the rest of low connected ones. This phenomenon leads to some relevant properties of the network, such as high diffusion rates of epidemics or attack resilience (Newman, 2010). Some variations of power-law distributions, which also include high clustering, have also been introduced to represent real networks (Dorogovtsev and Mendes, 2013).

Empirical analyses have found that some real SCs resemble this topological archetype. This is the case for some automotive SCs, which show a highly centralized network structure (Kim et al., 2011), and the supplier-customer network in the Indian auto-component industry (Parhi, 2005). The structure of relationships in these examples is represented by a heterogeneous degree distribution of links, more specifically a power-law degree distribution. This organization has also been identified in certain industrial districts, dominated by few leading firms which outsource production among a big number of smaller firms. This organization favors the adaptation to market changes from the leading to the subcontracted firms helped by production redundancy, proximity and long-term transactional intensity (Lazerson and Lorenzoni, 1999).

Nevertheless, other network topologies have been described in industrial districts as well: 1) Many small firms buying and selling material to each other; 2) Satellite platforms, formed by branches of large and external headquarter firms with few trade among them (Markusen, 1996). These patterns may be represented by a homogeneous degree distribution network with high or low clustering coefficient, respectively.

Likewise, other possible interaction patterns in real SCs have been studied recently (Capaldo and Giannoccaro, 2015a; Kim et al., 2015). Some of them have been identified in other fields, such as Poisson random networks (Newman, 2003), while other not so common topologies have been described as well. For example, the block-diagonal pattern is characterized by the existence of blocks of firms highly interconnected but with few connections among blocks. Other topologies are hierarchical, where a dominant firm includes many suppliers, or dependent, where few wholesalers include many independent retailers.

*2.4. Network topology of a resilient SC*

The interrelationship pattern of firms influences on the SC resilience. In particular, SC understanding is enhanced through the knowledge of the SC topology, which allows the identification of bottlenecks and critical paths in the supply network. Additionally, diversification of suppliers, which has been recommended as a measure to reduce disruption risks in SC (Sheffi and Rice, 2005), determines the network structure in the SC and their benefits can be also analyzed using topological tools.

Starting from findings in the previous section, some theoretical contributions hypothesizes that scale-free topologies characterize the most agile and resilient SC (Hearnshaw and Wilson, 2013). Similarly to the case of the industrial districts, SC resilience is achieved by the coordination role of the leading firm, which enhances communication and organization among suppliers, consequently improving the fast adaptation to changes. Leadership is also necessary to promote risk management culture, which increases resilience in SC (Christopher and Peck, 2004). Thus, hub firms promote horizontal links among suppliers, increasing clustering in the supply network. Therefore, high clustering is also a characteristic of a resilient SC, since it facilitates coordination among firms and thereby agility.

Some simulation models have also analyzed the SC resilience obtained with different topologies. For example, Thadakamalla et al., (2004) and Zhao et al. (2011) simulate several random network topologies in a three-tiered SC (supplier, manufacturer and retailer). In these papers, agility is measured through the average supply-path length, which is part of a collection of resilient indicators. The results find that heterogeneous structures, such as those induced by scale-free topology and variations, show the highest values of resilience when a random disruption occurs, but exhibit fragility when targeted attacks are produced. In general, these results agree with previous findings in a more general context (Newman, 2010).

Additionally, Nair and Vidal (2011) found that, in terms of inventory levels, scale-free topology performs better against disruptions than other homogeneous topologies but worst from the backorders and total cost perspective. They achieve the results by means of an agent-based model in a simple and reduced SC (only 35 nodes). Recently, Kim et al. (2015) analyze the resilience, measured as the probability of SC disruption in case of a node or link disruption, for a simulated supply network with several topologies, showing that the best figures are obtained with a power-law.

Some other simulation models have tested the SC performance with several topologies for other characteristics than resilience. Thus, Li et al. (2013) also found that heterogeneous structures also induce cooperation among firms in a greater extent than homogeneous structures. Additionally, Capaldo and Giannoccaro (2015b) show that a dependence pattern, followed by power-law, obtains the highest level of trust among partners.

*2.5. Agility in a SC with restricted relationships*

In general, the results above point to the prevalence of scale-free structures and derivatives as the most convenient topologies for a resilient and, in particular, agile SC. This is so for the types of SCs considered in the previous contributions. In general, the systems analyzed above are complex supply networks, where links between firms in any tier are possible. This conditions adapt to some real industrial SC (Kim et al., 2011), but not the FSC illustrated in section 2.1. In this kind of SC, a restricted relationship between agents exists, where every agent is supplied by other agents in the precedent tier and supply to agents in the subsequent tier. For example, a wholesaler is supplied exclusively by intermediaries, while an intermediary is only supplied by producers. These restrictions determine a specific supply network structure, which may influence on the SC agility. We call this structure *restricted relationship*. It is similar to the diagonal structure, where every firm is supplied from firms in upstream tiers (Capaldo and Giannoccaro, 2015a; Kim et al., 2015), but adding that firms are supplied only by firms in the precedent tier. Figure 1 illustrates two examples of this supply network structure in a three-tiered SC.

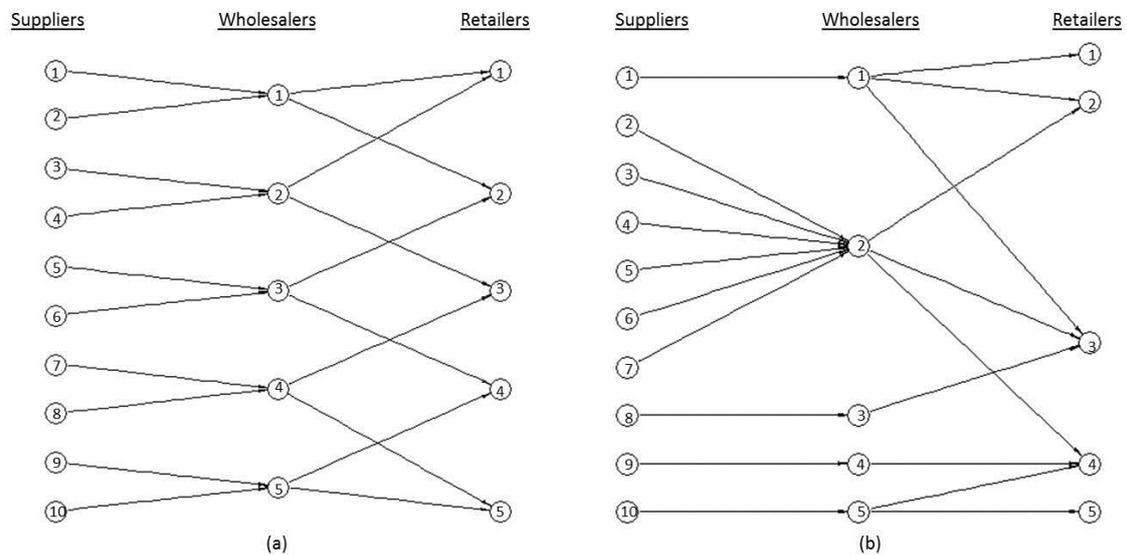

Figure 1. Two examples of a three-tiered supply network with different topologies in the wholesalers and retailers: (a) Homogeneous topology (regular degree distribution with mean degree $\bar{k} = 2$); (b) Heterogeneous topology (power-law degree distribution with $\bar{k} = 2$).

Thus, the type of SCs analyzed here does not reply those in previous papers, for which more flexible relationships among firms in the supply network are allowed. In a restricted relationship structure, the pattern of links from firms in a tier with firms in the precedent or subsequent tier determines the SC topology. For example, Figure 1a shows a realization of a regular degree distribution of links between supplier-wholesaler and wholesaler-retailer. This distribution assumes a constant degree ($\bar{k}$) for any firm in the SC. Figure 1b shows a realization for a scale-free degree distribution of links between the same tiers. The details for the construction of these SCs are given in the next section.

In the case of SCs with restricted relationship, scale-free degree distribution may not be the most suitable pattern to get the highest levels of agility. As shown in Figure 1b, the SC topology generated by this distribution presents few big firms that have many relationships (wholesaler 1 and 2) combined with small firms having few connections. Large differences between the number of suppliers and buyers of every firm are more likely presented with heterogeneous than with homogeneous distributions of links. For example, in Figure 1a all wholesalers and retailers have the same number of buyers and sellers, while in Figure 1b, wholesaler 1 has only one seller and three buyers. Consequently, assuming the absence of horizontal links among firms in the same tier, when sudden changes of demand occur, some sellers in a heterogeneous topology may be overloaded of demand (e.g. wholesaler 1 in Figure 1b) while other bigger sellers can fulfill their orders making use of part of their capacity (e.g. wholesaler 2). Therefore, the total orders in the SC would not be fulfilled quickly although it had enough goods from the suppliers.

This uneven response to demand is avoided with a homogeneous distribution of links among firms, such as the one illustrated in Figure 1a. In this case, the gap between the number of buyers and sellers in every firm is low or null, so the total unfulfilled orders could be reduced. From these arguments, we state the following hypothesis:

H1: In case of capacity constraints and absence of horizontal relationships, agility in a SC with restricted relationship is higher with homogeneous-degree topologies than with heterogeneous-degree topologies, such as those induced by the power-law distribution.

Horizontal relationships in the SC imply collaboration among competitors. This practice has been observed in food processing companies and is justified by the incentive of receiving back the help in times of crisis (Scholten and Schilder, 2015). The collaboration between competitors implies the transaction of information and material between partners when needed and enhances flexibility, since it assures new material or service to customers.

From the topological point of view, agility is related to path length in a supply network. In SCs with restricted relationship, *N* tiers and non-horizontal relationships, the path length from any supplier connected to the final demand is constant and equal to *N-1*, the number of links from the initial supplier to the final customer. If horizontal relationships exist, firms can buy and sell to other firms in the same tier when needed. Thus, other new alternative paths for the flow of material from the bottom supplier to the final customer are included, which help to a quick response when sudden increases in demand happen.

We posit the second hypothesis of this paper:

H2: The larger the number of horizontal relationships among firms in the same tier, the more agile a SC with restricted relationship is.

*2.6. Topological metrics of SC agility*

Although strongly recommended by researchers (Borgatti and Li, 2009), the use of topological metrics to analyze supply networks is quite limited by now. These metrics are based on graph theory and describe the characteristics of a network exclusively by the disposition of nodes and links in the supply chain. One of the relevant contributions was made by Kim et al. (2011), who apply node-level and network-level metrics to three real automotive supply networks, finding some results who complemented those obtained by Choi and Hong (2002) using a qualitative approach. Among the network-level metrics, the authors use supply network centralization, which measures how a supply network is centralized around a focal firm, and network complexity, which combines the observation of the number of nodes and links. Previously, Falasca et al. (2008) propose the use of spatial density, complexity and node criticality as determinants of SC resilience.

To date, a specific topological metric for agility in SC has not been proposed yet. In general, it uses to be embedded in a mix of metrics of resilience in a supply network. For example, Zhao et al. (2011) proposes three indicators of a resilient SC: high percentage of retailers having access to previous tiers, large size of the connected supply network and low average supply path length between any pair of supplier and retailer nodes. Other authors (Mari et al., 2015; Nair and Vidal, 2011) also include high clustering as a characteristic of a resilient SC. Among these topological metrics, agility may be specifically represented by the average supply path length. This measure indicates the average number of agents that the flow of material has to pass through before achieving the final user. Since every firm that manages the product includes some delays associated with receiving, processing and delivering, a low average supply path length means that the material passes through few hands, therefore reducing the total delay and increasing agility.

The metrics above do not take into account the positive effect of having alternative paths when needed. Additionally, they are determined exclusively by the static topological structure of the SC and do not consider the operational characteristics of the network. The SC is conceived to transport some flow of material or information from the first tier to the final assembler or consumer. This transportation takes also some time, which is essential for analyzing the response of the SC when a disruption occurs. So, the resilience metrics in general and agility metrics in particular should include the time factor in their specifications.

In this line, the percentage of satisfied demand over the total demand in a certain period of time, also called order fulfillment rate (OFR), include these aspects. This metric has been traditionally used as a quantitative SC performance measure (Beamon, 1998; Levy, 1995). More recently, some authors propose OFR as part of other resilience measures of SC as well (Barroso et al., 2015; Datta et al., 2007). In particular, Barroso et al. (2015) uses the concept of resilience triangle, which can be defined as the performance trajectory along time since a disruption occurs until the SC achieves the normal performance levels (Tierney and Bruneau, 2007). Other metrics based on the resilience triangle can be found in the paper of (Carvalho et al., 2012), which propose as part of the resilience metric the ratio of the actual and the promised lead time to deliver an order.

Alternatively, Plagányi et al. (2014) proposes the Supply Chain Index (SCI), a quantitative metric to identify resilience and connectedness of a SC. The novelty of this index is that it includes the flow of material throughout the SC in the definition. Low values of SCI indicate that the flow of material is diffused among multiple agents, while large values imply high concentration of flow in few nodes. The SCI was tested by several simple examples of linear SCs. The authors find that the most favorable values of SCI in terms of agility depend on the form of adaptation cost of changes. In general, diffuse flows favor quick adaptation to changes if the cost of changing flows from one agent to another is not scale dependent.

This paper adopts the operational perspective of the previous papers and measures agility through the OFR. Instead of analyzing the response of the SC when any class of disruption occurs, only sudden shocks in demand are assumed. Thus, a specific metric for agility is obtained. It is assumed that the lowest suppliers produce a certain amount of material in a specific period of time, which is transported downstream until the final customer. For simplicity, the simulation model does not include delivery times and inventory in every firm. Instead, flow of material throughout the SC is calculated through the demand from the final customer and the processing capacity of every firm.

## 3. A model to test supply chain agility

### 3.1. Supply chain random network

In this section, we present a stylized model for a supply chain, called supply chain random network (SCRN), where the response to demand shocks will be simulated. Figure 2 shows the general representation of the model. According to a topological viewpoint (e.g., Li et al., 2013; Tang et al., 2016), a supply chain is a directed random graph $G(V,E)$, where $V$ indicates the vector of tiers and $E$ the set of links among nodes (firms) in the graph. In general, a supply chain can include multiple tiers, but for the sake of simplicity we limit to three in this model. So,

$$V = (T_s, T_w, T_r),$$

where $T_h = \{h/h = 1,2,\dots,N_h\}$, with $h=\{s,w,r\}$, are the sets of nodes in the tier corresponding to supplier, wholesaler and retailer, respectively. These sets do not include any node in common and the number of nodes in every tier is not necessarily the same. We denote $\alpha:1:\beta$, with $\alpha$, $\beta$ positive scalars, the ratio of the number of nodes between subsequent tiers (e.g., if $N_w = 100$, notation 3:1:5 indicates that the number of suppliers, wholesalers and retailers are 300, 100 and 500, respectively).

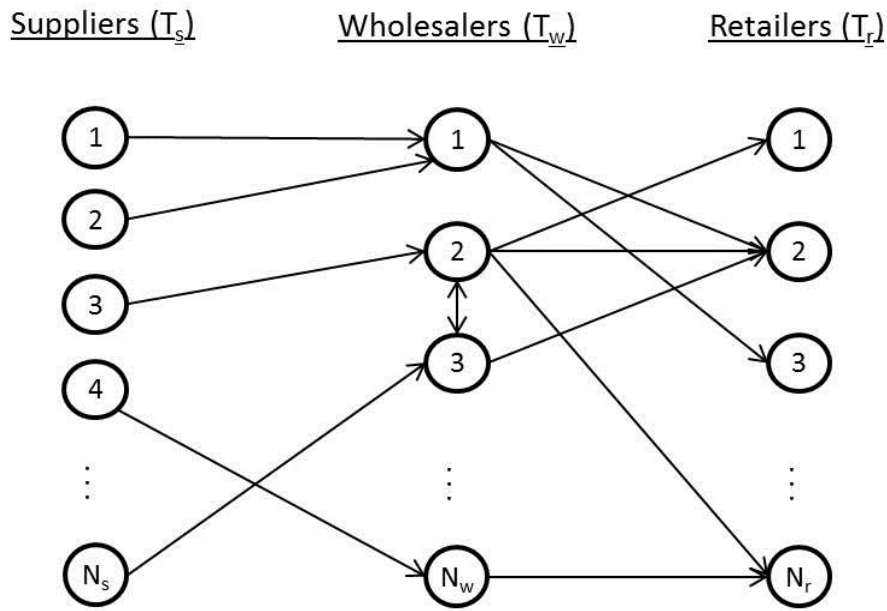

Figure 2. Representation of the supply chain random network with three tiers: Suppliers ($T_s$), Wholesalers ($T_w$) and Retailers ($T_r$). $N_s, N_w$ and $N_r$ represent the number of suppliers, wholesalers and retailers, respectively. Arrows indicate flow and direction of material between firms. The bi-directional arrow indicates horizontal relationship.

The links included in $E$ indicate the existence of a buyer-seller relationship between two firms, so a flow of some material (product) is transmitted between them when necessary and available. By assumption, links only connect nodes between subsequent tiers $T_s \rightarrow T_w \rightarrow T_r$, so a direct nexus $T_s \rightarrow T_r$ is not considered. Therefore, *E* includes pairs $(s,w)$ or $(w,r)$ if node *s* transmits the product to *w*, and node *w* transmits product to *r*, with *s*, *w*, *r* in $T_s, T_w, T_r$, respectively. Additionally, links between nodes in the same tier (horizontal relationships) are allowed only for wholesalers. The horizontal relationships are bi-directional, so the two firms in $T_w$ that are connected (2 and 3 in Figure 2) can send material to each other in order to satisfy their corresponding demands. In set-builder notation, this means that *E* also includes pairs $(w,w')$ and $(w',w)$, with $w, w'=1,2,\ldots,N_w$. For simplicity, the simulation model does not include delivery times and inventory levels in every firm. The analysis of agility is restricted to the SC's capability to fulfill instantaneously the orders coming from retailers.

Every node in $T_s$ and $T_w$ includes a weight (capacity), which indicates the maximum volume of product that the node can produce or distribute, respectively, in a certain unit of time. For simplicity, we assume that every producer has identical production capacity, $c_s$. By default, we also assume that every wholesaler has the ability to distribute the volume of product coming from his/her relationships with producers. So, the wholesaler's capacity is equal to the number of producers he/she has relations with (i.e. the wholesaler's in-degree) times the producer's capacity, $c_s$. Thus, in the example shown in Figure 2, wholesaler 1's capacity is $2c_s$, since it is linked with two suppliers, while wholesaler $N_w$'s capacity is $c_s$, since it is linked to only one supplier.

Finally, the topology of the supply chain is a combination of two bipartite random graphs (Newman and Watts, 2002). They are the supplier/wholesaler and wholesaler/retailer random graphs. Every supplier *s* has $k_s^{out}$ links to the wholesalers and every retailer *r* has $k_r^{in}$ links from the wholesalers. Every wholesaler *w* has $k_w^{in}$ links from the supplier and $k_w^{out}$ links to the retailers. We assume that the degree distribution among nodes in every tier (two in-degrees and two out-degrees) follows some prescribed probability distributions. A sample of the random graph is obtained by following the same procedure used for the configuration model (Guillaume and Latapy, 2006): First, we generate the in- and out-degree of every node by simulating the specific distribution and every node is assigned a number of "stubs" equal to its degree; Second, we connect the stubs between subsequent tiers randomly, creating links among the nodes. In order to assure consistence (the sums of the degrees in subsequent tiers must be identical), the mean of the distribution times the number of vertices in subsequent tiers must be identical. Nevertheless, the specific sample may be inconsistent yet. In this case, one node from each one of the two tiers is dropped and their degrees are redrawn until the sample is consistent.

*3.2. Model of product distribution*

We assume a determined system of product distribution and order allocation in the SCRN. This procedure will allow obtaining a measure of the OFR, which depends exclusively on every node's capacity and the network topology.

Previously, the total demand in the system is assumed to be identical to the total production capacity of suppliers. Therefore, the material demanded coincides with the disposable supply and, in the naïve supply network with just one supplier, wholesaler and retailer, all demand would be satisfied. However, this may be not the case when several firms operate in every tier, such as it is the case in the SCRN.

First, we assign the order allocation rule. It is assumed that every retailer orders one unit of material for every link that the retailer maintains with the upstream tier. In other words, this means that the retailer decides the number of relationships with wholesalers according his/her necessities. In mathematical terms, the retailer *r* orders $D_r = k_r^{in}$ units, with $r=1,2,\ldots,N_r$, and the total demand from $T_r$ is $D = \sum_{r=1}^{N_r} k_r^{in} \approx \bar{k}_r^{in} N_r$, with $\bar{k}_r^{in}$ the mean value of the in-degree distribution $k_r^{in}$. Each supplier's capacity, $c_s$, is determined to assure that, in mean terms, total supply is equal to total demand, so

$$c_s N_s = \bar{k}_r^{in} N_r. \qquad (1)$$

Since $N_s = \alpha N_w$ and $N_r = \beta N_w$, we have that $c_s = \beta/\alpha \, \bar{k}_r^{in}$.

Second, we define the supply allocation rule followed by wholesalers. Let us denote $e$ the accumulated unfulfilled orders, which is initially set as $e=0$. Starting from the retailer $r=1$, let $w_r$ be the wholesaler linked to $r$ with the lowest position in $T_w$. In other words, if $w_r'$ is other wholesaler linked to $r$, then $w_r < w_r'$. As stated in the previous subsection, $w_r$'s initial capacity ($c_{w_r}$) is equal to $w_r$'s in-degree times each supplier's capacity, so $c_{w_r} = k_{w_r}^{in} c_s$. Starting from $c_{w_r}$ and $D_r$, two cases are possible:

1. $c_{w_r} \geq D_r$. This means that the wholesaler's capacity exceeds the $r$'s demand, so the quantity $D_r$ is moved from $w_r$ to $r$ and $w_r$'s capacity reduces to $c_{w_r} - D_r$.

2. $c_{w_r} < D_r$. In this case, all the material distributed by $w_r$ is sold to $r$. So, an amount $c_{w_r}$ is moved from $w_r$ to $r$ and $w_r$'s capacity transforms to $c_{w_r} = 0$. Then the following $w_r'$ (in numerical order) that links $r$ is considered and the same procedure is applied previously re-assigning $r$'s demand $D_r' = D_r - c_{w_r}$. The process is repeated recursively until achieving case 1 or $c_w < D_r'^{\cdots\prime}$ for the largest $w$ that links $r$. In the latter case, a quantity $c_w$ is moved from $w$ to $r$ and the unfulfilled orders $e$ increases in $D_r'^{\cdots\prime} - c_w$.

Once the orders coming from $r$ are completely managed, the process is repeated for retailer $r+1$'s and so on until reaching $r=N_r$.

The model resembles the real transmission of material in a supply network, although it includes several strong restrictions. First, identical volume of transactions is assumed for every buyer-supplier relationship. Second, the supply allocation rule assumes a kind of hierarchy among retailers, since any retailer $r$ has preference to be satisfied over all $r'>r$. Third, any wholesaler must satisfy all the demand from a particular retailer before attending the next retailer in the hierarchy. Other SC simulation models have included alternative order and shipment allocation rules, such as maintaining inventory levels (Datta et al., 2007; Levy, 1995) or assigning preference based on backorders and trust measures (Kim, 2009). These factors are not considered in the model. Any kind of strategic behavior of agents is not included either.

The algorithm initiates with $r=1$ and covers all retailers subsequently. The final value of $e$ indicates the amount of unfulfilled orders. The ratio OFR=1-$e$/D gives the percentage of fulfilled orders from the total demand or order fulfillment rate. The larger the value of OFR in [0,1], the higher the agility of the supply network.

### 4. Simulation setup and results

In order to test the influence of the network topology on the SC agility, we simulate the performance of the model above considering several degree distributions of firms in a tier. These degree distributions are the result of the specific conditions in the market and each individual's decisions to have relations with some of the downstream or upstream partners. Up to ten possible supply network structures have been recently identified in the literature (Capaldo and Giannoccaro, 2015a), which have been reduced to four by Kim et al. (2015). Nevertheless, since our main aim is to check whether heterogeneous degree distributions, such as power-law, favor or hinder SC agility to a larger extent than other homogeneous degree distributions, we will limit the analysis to some stochastic distributions which represent the alternatives we look for comparison.

Thus, three discrete degree distributions have been selected in the simulations: Regular, Poisson and power-law. A regular degree distribution assumes that every firm in a tier has exactly the same number of relationships with other downstream or upstream firms. A Poisson degree distribution allows the firms to have different number of relationships, but around a mean value. Assuming these two degree distributions, the characteristics of the firms in terms of flow of material and transactions are similar and there is not a leading position of a firm with respect to another. Both are homogeneous distributions, since firms in this kind of SC have proximately similar position in terms of number of relationships. On the contrary, a power-law distribution assumes sharp differences in firms' position, since it includes a short but significant number of firms accumulating many of the total relationships, while the rest has few links with the other tier.

All distributions in the simulations are zero-truncated, so every node in the network has at least one link with a firm in the upper/lower tier. By means of this restriction, we set aside the case of suppliers or retailers which are not integrated in the supply chain or wholesalers without relationships with either producers or retailers. In order to compare coherently the performance of the sampled supply networks, identical mean value is assigned for the three distributions.

Based on a recent estimation of a real-world food supply chain, in particular the seafood supply chain in Guadalajara, Mexico (Pedroza and Hernández, 2016), we have considered a data set of 1300 firms distributed according the ratio 2:1:10, so 200 suppliers, 100 wholesalers and 1000 retailers. The SCRN and the algorithm to calculate OFR was implemented in Matlab®. As described in section 3, we assume that every retailer $r$ demands $k_r^{in}$ units of material. Therefore, the total demand is approximately $D \approx 1000\bar{k}_r^{in}$. Using the relationship described in (1), each supplier produces $c_s = 5\bar{k}_r^{in}$ units. We consider that wholesalers have the same behavior in their relations with suppliers and retailers, so the wholesaler's in- and out-degree distributions are assumed identical. Given these conditions, the possible combination of the three distributions among three tiers is still $3^3$=27. In order to simplify the results, we also assume that every supplier trades with only one wholesaler, so the out-degree distribution $k_s^{out}$ is regular with mean degree equal 1. By doing this, we limit the scope of the analysis to the comparison of SC agility obtained with three topological structures in two tiers (wholesalers and retailers), which are $3^2$=9.

Table 1 presents the results of the OFR for the three degree distributions included in the analysis. Several values of mean in-degree for retailers $\bar{k}_r^{in}$ is assumed. As it is shown in the four cases presented, in general the highest agility of the supply network is achieved when degree distributions are homogeneous (regular or Poisson). Lower agility results are obtained when assuming power-law distributions. Nevertheless, the figures also show that SC agility depends on the combination of specific probability laws followed in one or another tier. Thus, given that retailers adopt a heterogeneous distribution of links, the SC agility heavily decreases if the wholesalers follow regular or Poisson distributions, while recovers if they also follow a power-law distribution. However, this is not so if a heterogeneous distribution of links in wholesalers is combined with a homogeneous distribution at the level of retailers. In this case, the agility indicators are not far from those achieved with homogeneous distributions in both tiers.

Table 1. Order fulfillment rate (OFR) in the supply chain random network (SCRN). Every supplier includes only one link with wholesalers. Three discrete distribution functions for the wholesaler's in- and out-degree ($k_w^{in}$ and $k_w^{out}$) and the retailers' in-degree ($k_r^{in}$) are considered: Regular (Reg.), zero-truncated Poisson (Poiss.) and zero-truncated power-law (Pow.). The ratio among the number of firms in tiers is 2:1:10, the number of wholesalers is $N_w = 100$ and no horizontal relationships among wholesalers exist. The cases are, from the left to the right: (a) The mean in-degree of retailers is $\bar{k}_r^{in} = 2$ and relationships among wholesalers and retailers are randomly assigned, not having into account in and out-degrees; (b) $\bar{k}_r^{in} = 2$ and relationships among wholesalers and retailers are ordered according to their degrees, so the retailer with the highest number of relationships trade with the wholesaler with the highest out-degree, and so on; (c) Same as (a) with $\bar{k}_r^{in} = 4$; (d) Same as (a) with $\bar{k}_r^{in} = 8$.

| | $k_r^{in}$ | | | | | | | | | | | |
|---|---|---|---|---|---|---|---|---|---|---|---|---|
| | (a) $\bar{k}_r^{in} = 2$ | | | (b) $\bar{k}_r^{in} = 2$ (ordered) | | | (c) $\bar{k}_r^{in} = 4$ | | | (d) $\bar{k}_r^{in} = 8$ | | |
| $k_w^{out*}$ | Reg. | Poiss. | Pow. | Reg. | Poiss. | Pow. | Reg. | Poiss. | Pow. | Reg. | Poiss. | Pow. |
| Reg. | 0,90** | 0,85 | 0,52 | 0,90 | 0,85 | 0,53 | 0,93 | 0,89 | 0,53 | 0,96 | 0,93 | 0,53 |
| Poiss. | 0,90 | 0,85 | 0,56 | 0,95 | 0,91 | 0,57 | 0,93 | 0,89 | 0,56 | 0,96 | 0,93 | 0,55 |
| Pow. | 0,87 | 0,83 | 0,71 | 1 | 0,99 | 0,72 | 0,90 | 0,87 | 0,84 | 0,94 | 0,92 | 0,92 |

*Identical probability distribution law for the wholesaler's in- and out-degree is assumed. To assure consistency, the mean degree $\bar{k}_w^{out} = \beta \bar{k}_r^{in} = 10 \bar{k}_r^{in}$ and $\bar{k}_w^{in} = \alpha \bar{k}_s^{out} = 2$. The sample of $k_w^{in}$ and $k_w^{out}$ is ordered, so the wholesaler with the highest in-degree has also the highest out-degree, and so on. All distributions are zero-truncated.

**The figures are obtained by doing 1000 simulations of the SCRN and taking mean values. In order to avoid high differences between total demand and supply in the simulations, samples with a gap between the simulated and theoretical mean degree higher than 5% are not considered.

Therefore, given the conditions of the SCRN, the power-law distributions are not the most efficient degree distribution in the supply chain in terms of fast response of supply changes. This is particularly marked when the bottom tier (retailers) follows the power-law distribution of links. These results confirm H1 above and simulations go in the line of reasons given when stating the hypothesis: The uneven demand from retailers lead that some retailers consume a big part or the total capacity of many small wholesalers, implying that some of the other retailers also linked to those wholesalers cannot satisfy their necessities. This effect is not so marked if there are not such big leading retailers, such as it is the case when retailers follow homogeneous degree distributions.

The depletion of many wholesalers' capacity by the leading retailers could be partially avoided if these retailers trade also with leading wholesalers. This possibility is captured in Table1b, which assumes a positive correlation of the wholesaler and retailers' degree, what is also called positive assortativity. This condition means that retailers with high demand tend to relate with leading wholesalers with high number of relationships. The results for agility are something higher but do not differ very much from those obtained in the non-correlated case (Table 1a). The effect of positive assortativity between wholesalers and retailers is only significant for the SC agility when wholesalers follow a power-law degree distribution and retailers adopt homogenous degree distributions.

Tables 1c and 1d show the results of OFR assuming higher mean in-degree of retailers ($\bar{k}_r^{in} = 4$ and $\bar{k}_r^{in} = 8$, respectively). In order to assure consistency, the mean out-degree of wholesalers and production of suppliers increase consequently. By increasing the number of relationships among wholesalers and retailers, we analyze the scale effect on agility. The results show again that the SCRN performs better when assuming homogeneous distributions of links between all tiers than mixing homogeneous and heterogeneous distributions. However, when both tiers follow power-law distributions, the agility is practically identical to consider Poisson and regular degree distributions. Thus, power-law distributions can be efficient when high number of relationships is presented. This result agrees with the common redundancy recommendation of having multiple suppliers to increase SC agility (Sheffi and Rice, 2005), but restricting it to the case of supply networks with heterogeneous distribution of links.

Horizontal links among wholesalers may enhance agility, since the related firms interchange production between them when needed to fulfill their own orders. Figure 3 shows the OFR evolution when increasing the percentage of the total wholesalers with horizontal links ($\rho$). In the simulations we consider that two linked wholesalers function as one, adding their capacities and orders. The case $\rho$=0 presents the same conditions and gives the same OFR values shown in Table 1a, while $\rho$=1 results in the maximum agility (OFR=1) for any degree distribution since, in this case, all wholesalers share their product and therefore the supply and demand are perfectly connected. In between, OFR increases between the two extreme values, but not regularly. It is not very much affected when less than half of wholesalers interchange production and the steeper increase is produced when horizontal links are higher than 80%. The effect of increasing horizontal links is negligible up to point when homogeneous in-degree distributions of retailers are assumed. In the case of power-law in-degree distribution, the positive effect on agility starts to be intense when more than 50% of wholesalers are linked. These results confirm H2, but provide additional information of the benefits of having horizontal links. The marginal positive effect on agility of adopting new horizontal links in the SCRN is low when starting from low interrelationships between wholesalers but high when the number horizontal links exceed a certain threshold.

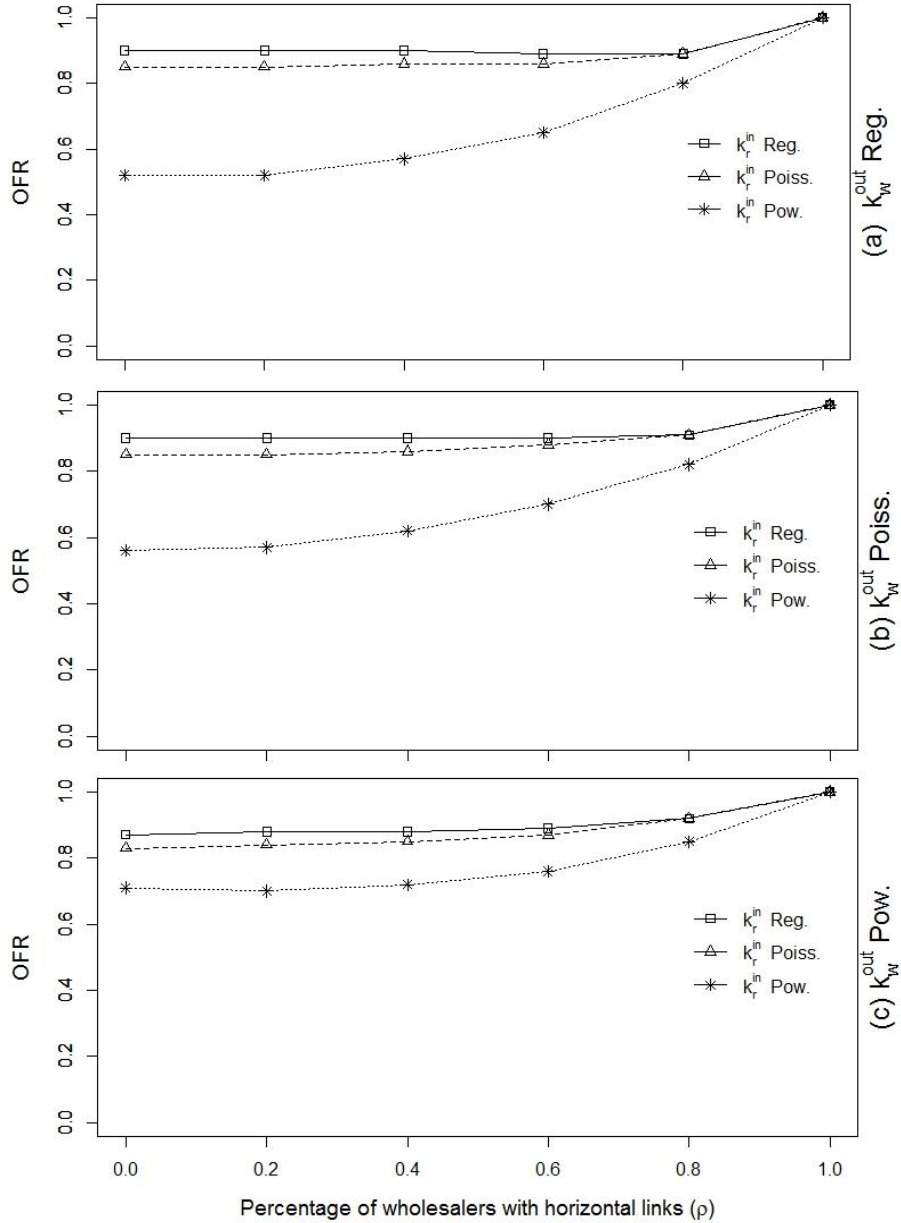

Figure 3. OFR with respect to the percentage of wholesalers with horizontal links (*ρ*). The initial conditions are identical to case (a) in Table 1. The three trajectories in every graph correspond to three in-degree distributions for retailers $k_r^{in}$: Regular (Reg.), zero-truncated Poisson (Poiss.) and zero-truncated power-law (Pow.). Every graph assumes the following out-degree distributions of wholesalers ($k_w^{out}$): (a) Regular; (b) Zero-truncated Poisson; (c) Zero-truncated power-law. Every OFR value is obtained by doing 1000 simulations of the SCRN and taking mean values.

## 5. Discussion

This paper aims to analyze the influence of the topology on the SC agility. This is done by building an agent-based model of a SC including three tiers and some other specifications, such as restricted relationship between firms. The latter condition does not adjust to some industrial supply chains, such as automotive (Choi and Hong, 2002), but does to food supply chains which are compounded by small-scale producers located in rural development economies and unable to take their products to outside markets (Abebe et al., 2016; Crona et al., 2010; Hayami et al., 1999; Merlijn, 1989; Pedroza, 2013). In these circumstances, direct trading between wholesalers and producers is mostly absent, so middlemen exercise a coordination role facilitating product transmission. Abebe et al. (2016) describes this type of food supply chains as "where market failure is ubiquitous and food chains still consist of many stages" and where middlemen represent a socially-tied relationship.

Agility, or quick response to sudden demand changes, is treated as a dimension of the SC resilience, understood here as the ability of the SC to respond to disruptions and disturbances. Contrary to some previous experimental contributions which found that scale-free topologies characterize the most resilient SCs (Kim et al., 2015; Nair and Vidal, 2011; Zhao et al., 2011), the simulation results show that this is not so for any kind of supply networks. Specifically, in the SC conditions analyzed in the paper, homogeneous topologies, such as those which follow regular or Poisson degree distributions, assure in general higher level of agility than heterogeneous topologies, such as those derived from power-law degree distributions.

In previous contributions, some of the positive characteristics found in heterogeneous distributions for SC resilience are founded on the translation of common properties of complex networks (small world effect, high clustering) to the context of SC, such as it has been evidenced in some theoretical developments (Hearnshaw and Wilson, 2013). The design of the simulation experiments to analyze SC resilience, such as the presence or random disruptions and targeted attacks, also resembles those applied for complex networks in other fields, such as Epidemiology and Cybernetics (Newman, 2010), and the results obtained are fairly similar. Our results extend these previous findings by designing a simulation experiment conceived for SC and illustrate that the most suitable topology to achieve the highest level of agility, and consequently resilience, is not scale-free in the case of restricted relationship among firms.

Therefore, this study finds that a SC with similar number of relationships among buyers and sellers is under certain conditions more agile than a SC where few leading firms with many relationships are combined with many small firms with few relationships. This result does not agree either with previous theoretical developments in the context of SC resilience. Specifically, they argue that the presence of leading firms is beneficial to SC resilience by enhancing risk management culture and horizontal links among suppliers (Christopher and Peck, 2004; Kamalahmadi and Parast, 2016). However, in addition to these effects, the pursuit of leadership causes competition among firms in the same tier and likely the emergence of a topology with an uneven distribution of links. The findings in this paper show that this is not the most favorable structure for the agility of a certain type of SC. Which one of both opposite effects predominates is a matter of further empirical research and theoretical analysis. Looking at our findings exclusively, there is a trade-off between each firm' objectives, which are dominated by the maximization of revenues and the quick response to own orders, and the agility of the SC as a whole.

In addition to the role of leadership, redundancy (increase of relationships) has been also recommended to enhance SC agility and resiliency (Jüttner and Maklan, 2011; Sheffi and Rice, 2005). This view has been questioned by the resilience results obtained from a SC simulation model by Kim et al. (2015). In their model, resilience is measured through the prevalence of the SC functioning when node or link disruptions occur. On the contrary, our results agree with the common view in the literature and redundancy shows beneficial for SC agility. This effect is more pronounced when heterogeneous degree distributions, such as power-law, are presented in the SC. From the comparison of our results with Kim et al. (2015)'s, it can be concluded that in some types of SCs redundancy favors some dimensions of resilience, such as fast response to demand changes, while not having a considerable effect in other types of SC and dimensions, such as the adaptation to node or link disruptions.

The agility metric used in this paper (OFR) differs from the one used in other contributions (average supply path length). One of the reasons is the specific type of SC analyzed. The findings obtained in this paper are applied to a SC with a restricted structure of relationships, such as food supply chains, where each firm trades exclusively with partners located in the previous or next tier. If firms could relate each other freely, such as is the case in other industrial supply chains and commonly assumed in SC simulation models, closed triads and high clustering arise. Agility can be approached in this kind of SC by the average supply path length (Nair and Vidal, 2011; Zhao et al., 2011). Using this metric, the best agility performance was obtained with scale-free topologies. However, supply path length is not useful for SCs with restricted relationships, since the shortest path length from the first producer to the customer is constant, independently of the specific topology. Order fulfillment rate is a more suitable metric for agility in this kind of structures and the results show that in general SCs with scale-free topology do not reach the highest values of OFR. The different metrics used for each structure do not allow making a rigorous comparison of findings, but illustrates the opposite performance of scale-free topologies with respect to SC agility.

According to the previous analysis, it may be guessed that, in terms of agility, homogeneous distribution of links among nodes is preferable for food supply chains, where restricted relationship SCs are common. On the contrary, heterogeneous distributions or scale-free topologies are more recommendable for industrial supply chains, where relationships among firms are less restricted. However, the empirical analyses of real SCs in both food and industrial supply chains are very scarce up to date to classify them strictly in one or another type, so a conclusive assertion on this point cannot be given yet.

Moreover, it cannot be deduced from the findings that a resilient SC is not compatible with scale-free topology. The results are limited not only to a specific SC structure, but also to a specific dimension of resilience (agility). As it is has been shown in previous studies, other dimensions of resilience, such as the response to node/link disruption, are more favored with scale-free than with other homogeneous topologies (Kim et al., 2015). Other relevant properties of the SC are also enhanced by power-law distributions. This is the case of cooperation among firms (Li et al., 2013) and trust among partners (Capaldo and Giannoccaro, 2015b).

## 6. Conclusions, limitations and implications

This paper analyzes the influence of the pattern of relationships on the agility of the supply chain. SC agility means quick response to demand changes and is one of the characteristics of an efficient SC. Many previous theoretical and empirical contributions highlight that most efficient SCs are those with heterogeneous distribution of links, such as the one induced by scale-free topologies. By simulating an agent-based model built to represent a supply network and the flow of material, this paper shows that this is not the case for SCs where relationships are restricted to firms located in subsequent tiers. This type of SCs is presented in military logistic networks and some food supply chains. For a SC like this, homogeneous distribution of links results in higher agility than heterogeneous distributions of links, such as that one induced by power-law distribution.

Therefore, the results of this paper illustrate that the most efficient topology in a SC is not necessarily scale-free, but depends on the conditions of the specific SC and the specific aspect of efficiency. Nevertheless, the results confirm other common recommendations for resilient SCs, such as redundancy. Higher agility levels are obtained by increasing the number of relationships among firms in two subsequent tiers. Horizontal links between firms in the same tier also favor the SC agility, although their effect is significant when a certain volume of relationships is surpassed.

The methodological approach followed in this paper makes that findings must be taken with caution. SC agility is influenced by qualitative factors that are not included in the simulation model. To give some examples, visibility and the existence of leading firms, which favor SC agility, are not analyzed here.

Additionally, other limitations are presented. First, the algorithm includes a mechanistic way of trading, such as hierarchy in the order and supply allocation rules. Likely, this may not represent the real behavior of agents, which can follow some kind of strategy in the allocation rules. Second, it is assumed identical amount of product for every link, not allowing the firms to increase orders to the same seller. Third, it is not solved the optimal design of a SC with respect to agility. Power-law has been detected as the most suitable degree distribution among a selection of three, but it cannot be derived from the results that the induced topology from power-law distribution gives the best performance among any possible topology. Fourth, the analysis does not consider other strategies to improve SC agility, such as adding capacity and increasing inventory levels, which are considered constant in this paper.

Having in mind the limitations of the analysis, some managerial recommendations can be derived from the findings. They are circumscribed to those SCs in the conditions analyzed in the paper, i.e. restricted relationships among firms, where food supply chains are a suitable example. High levels of agility are specifically desirable in these types of supply chains, since they manage high volumes of perishable products that need fast distribution. According to the results, actions to avoid trade concentration in few agents belonging to the same tier (middlemen or wholesalers) are justified. By doing so, homogeneous distribution of links among agents in different tiers is favored, which enhances SC agility. Promotion of cooperation among agents in the same tier (horizontal links) is also recommended, although it is not expected that the effect on agility is sensible when cooperation is low.

This study can be extended in several ways. First, findings on agility here can be compared with new simulation results for general supply chain random network, removing the condition of restricted relationships among firms located in subsequent tiers. Additionally, this paper does not consider differences in the relationships among partners. However, strong and weak ties are present in SC relationships and show the degree of trade frequency or compromise between firms. The influence of the number and disposition of these ties on SC agility (or resilience in general) is still to be analyzed.

## Acknowledgments

The work was financed by project PAPIIT IA300215 from the Universidad Nacional Autónoma de México (UNAM) and project ECO2014-59067-P from the Ministry of Economy and Competitiveness of the Government of Spain.